\documentclass[12pt]{article}
\usepackage{graphicx}

\usepackage{latexsym}

\begin{document}
\title{On Controversies in Special Relativity}

\author{Anatoli Andrei Vankov,\\        
{\small \it Bethany College, KS; anatolivankov@hotmail.com}} 


\date{}

\maketitle

\begin{abstract}

Controversies arise when specialists disagree on some particular issue. This normally occurs in any scientific brunch. We analyze some controversies, which have a good cause in Special Relativity. The paper does not question Special Relativity Theory but it challenges changing deeply held academic
beliefs about imaginary experiments in Special Relativity Theory.  

 
\end{abstract}

\section{Introduction}

Special Relativity Kinematics is an old physical discipline, which is self-consistent and complete (what cannot be said about SR Dynamics), so it is strange enough that controversies arise in SR Kinematics. Our point is that they could be results of individual confusions, nevertheless, episodically appearing in media but hardly noticed or ignored if SR refutation or principal changes claimed. Sometimes confusions and controversies are caused by ambiguous or even wrong statements in reputable SR scientific literature and textbooks. In this work, we discuss some of such cases illustrating a historical lack of rigor in some presentations of SR concepts. (However, a systematic review of relevant publications is not our objective). The following issues are suggested for disputes.

-  Lorentz-to-Galilean transformation reduction (LT physical treatment).

-  Shape of light front from a moving source (sphere versus ellipsoid).

-  Relativity of simultaneity (prerelativistic versus relativistic concept).

-  Does a moving clock run slower? (proper versus improper time).

-  Main dogma in SR Dynamics (proper mass treatment).


\section{Lorentz-to-Galilean transformation reduction}
In recent work {\cite{Baierlein}, it is stated that, contrarily to a common belief (or ``myth''), ``the Lorentz transformation does not reduce to the Galilean transformation when the ratio $\beta=v/c$ is small''.  One needs to clarify this issue bearing in mind that typical textbooks present a formal (mathematical) rather than physical meaning of Lorentz transformations (LT).
Let us consider the special Lorentz transformation in the standard geometrical configuration ($\beta$-boost for a motion along $x$, $x'$ parallel axes without rotation of a spatial part in a Cartesian coordinate form). When applied to a 4-coordinate vector $X^\mu=(ct,\ x,\ 0,\ 0)$ for a free particle in a coordinate system $S$, it gives a vector $X'^\mu=(ct',\ x',\ 0,\ 0)$  in $S'$   
\begin{equation}
c t'=\gamma (c t -\beta x), \  \ x'= \gamma ( x -\beta c t), \  \  y'=0, \  \ z'=0
\label{1}
\end{equation}
It is argued that $ x$ in the first equation in (\ref{1}) (for a temporal component) can be however great at small $\beta$; hence, generally $t'\ne t$. This argument is readily supported, for example, by references Peres \cite{Peres} (p.249) and Sartori \cite{Sartori} ( p.102), but it is false: the equation in question is equivalent to $t'=\gamma t -v x/c^2$ and has the Galilean limit $t'=t$ at $c\to\infty$. Next, we summarize physical basics of the LT.

The Galilean-Newtonian Kinematics describes a particle motion in 3-space with a time $t$ playing a role of a common parameter. In general, Galilean coordinate transformations present one-to-one correspondence (3-space mapping) for every instant $ t $: $[x'(t),\ y'(t),\ z'(t)] \Longleftrightarrow [x(t),\ y(t),\ z(t)]$, the speed of light signal being assumed infinite.  Given an initial condition $x'=x=0$ at $t=0$, we have equations of motion of coordinate origins $x'(t)=-vt$ in $S'$ or $x(t)=vt$ in $S$, respectively. As for SRT Kinematics, it is based on the LT, perception of which is hard for Euclidean mind in spite of analogy with Galilean transformation. The LT is an expression of the Galilean relativity principle in a form of 4-space mapping  
$(x',\ y',\ z',\ t')  \longleftrightarrow (x,\ y,\ z,\ t) $, the postulate of speed of light constancy being added. The $t$ is not a parameter there but the 4-th (temporal) coordinate which can be mixed with a spatial part of the 4-coordinate vector. Tails of 4-vectors are put in a common origin of coordinate systems: $x_0=x'_0=0$ at $t_0=t'_0=0$ (a null, or $0$-event) chosen arbitrarily. Tips of vectors point at some another event represented by a 4-point $P$ in $S$ or $P'$ in $S'$ in Minkowski space.  
The analogous equations of motion for coordinate origins are  $x'=-vt'$, $x=vt$. It is seen again that in a Galilean limit the term $\beta x$ in  (\ref{1}) being of a second order in $\beta$
should be zeroed giving $\gamma=1$ and $t'=t$.

The $0$-event is a part of the LT concept and all its applications. So far, it is not clear how to realize it operationally in consistence with the postulate of speed of light constancy. Before deliberating this problem, we shall learn how to treat the LT in methodology of imaginary observers (say, John in $S'$ and Mary in $S$) both conducting metric measurements with the use of standard clocks and rods. 
Because of LT linearity, one can use space and time coordinate intervals $\Delta x=x-x_0$, $\Delta t=t-t_0$ rather than coordinates.  
In (\ref{1}), time intervals $\Delta t'$ and $\Delta t$ measured by John and Mary, correspondingly, are the so-called improper intervals (for their measurements more than one clock are needed). For better understanding a physical meaning of LT, let us see how a difference between proper and improper quantities vanishes in the Galilean limit. To find a relationship between improper and proper quantities, we need to specify observational conditions. 

Let us introduce proper time intervals  $\Delta t'_0$ and $\Delta t_0$  measured by each observer with the use of his/her individual wristwatch (a standard clock). Those intervals are, of course, equal. They are temporal components of proper vectors describing an observer's state of rest  ${\Delta X_0^\prime}^\mu=(c\Delta t'_0,\ 0,\ 0,\ 0)$  and   ${\Delta X_0}^\mu=(c\Delta t_0,\ 0,\ 0,\ 0)$. Suppose, Mary wants to conduct a distant measurement of John's time rate. She can do it using a method of signal (information) exchange: John's clock should work as an emitter of light signal, and Mary's clock as a corresponding detector. Her set of {\it improper} observational data are put in correspondence with the set of John's {\it proper} data by means of the inverse LT, the latter being applied to the proper (primed) vector:
\begin{equation}
\Delta x=\gamma \beta c\Delta t'_0, \  \  c \Delta t=\gamma c \Delta t'_0 
\label{2}
\end{equation}
It gives $X^\mu=\gamma\Delta t_0 (1,\ \beta,\ 0, \  0)$, where 
\begin{equation}
\gamma\Delta t'_0=\Delta t 
\label{3}
\end{equation}
This is a time dilation effect, which is a kinematical relativistic effect. Of course, John and Mary can interchange their roles in information exchange experiment: John carries a clock-detector to make observations at distance while Mary carries a clock-emitter. Again one has to apply the LT, this time the direct one, to the Mary's proper set of data (the 4-vector)  $\Delta X_0^\mu=(c\Delta t_0,\ 0,\ 0,\ 0)$. The result is: 
\begin{equation}
\Delta x'=-\gamma \beta c\Delta t_0, \  \  \gamma \Delta t_0= \Delta t' 
\label{4}
\end{equation}
The Galilean limit immediately takes place at $\gamma\to 1$.

The improper length is introduced when a distance of travel of one frame with respect to another is specified with the use of standard measuring rods. For example, Mary can measure John's time of travel from $x=0$ to the point $x=\Delta x_0=c\beta\Delta t$. Here, $\Delta x_0$ is the proper length interval and $\Delta t$ is the improper time interval in Mary's reference frame. In another experiment, she can measure her proper time interval $\Delta t_0$ between the $0$-event and the event of coincidence of her origin $O$ with John's mark $x'=-\Delta x'_0$. She measures a corresponding improper length interval $\Delta x$ such that 
$\Delta x=c \beta \Delta t_0$. Of course, the observers can interchange their roles. In accordance with the Galilean principle of relativity, we have following relations between proper and improper quantities arising in direct and inverse LT: 
\begin{equation}
c \beta=\Delta x_0/\Delta t=\Delta x/\Delta t_0=\Delta x'_0/\Delta t'=\Delta x'/\Delta t'_0
\label{5}
\end{equation}
where primed and unprimed quantities relate to John and Mary, respectively, and a zero subscript denotes properness. With ({3}) taken into account, we have the length contraction effect     
\begin{equation}
\gamma\Delta x=\Delta x_0 
\label{6}
\end{equation}

The LT exploits a mathematical identity $\gamma^2-\gamma^2\beta^2=1$ of hyperbolic rotation ensuring invariance of
the vector length (norm) in Minkowski space. The vector norm is the proper time (or time interval). 
In general, a time-like quadratic metric form (the squared norm) is invariant under LT:
\begin{equation}
(ct)^2-(x^2+y^2+z^2)=(ct')^2-(x'^2+y'^2+z'^2) 
\label{7}
\end{equation}
From this, the time dilation follows as a relative relativistic effect
\begin{equation}
t^2(1-\beta^2)=t'^2(1-\beta^2)={t'_0}^2= t_0^2 
\label{8}
\end{equation}
If interpreted physically, it says that Mary and John agree that their standard clocks run synchronously locally in all reference frames (in terms of proper time). When measured at distance, an elapsed (improper) time becomes dependent on a relative speed. The difference between proper and improper time does not mean a preference for any observer in inertial motion. Clearly, in Galilean limit the difference between proper and improper quantities vanishes, and the LT naturally reduces to the Galilean transformation. The latter automatically preserves the time interval due to the absolute character of time.

\section{Shape of light front, and constancy of the speed of light}
\subsection{Typical textbook presentations} 

There are interesting historical facts concerning the postulate of constancy of the speed of light, Baierlein\cite{Baierlein}. 
Putting details of the history aside, we want to discuss the postulate in connection with a relativistic picture of light front observed in different inertial reference frames provided photons are emitted from a single point source. Baierlein's view on this matter is in concordance with reputable SR literature; however, there are controversies about it, as discussed next.   
We again consider a standard configuration of two (primed and unprimed) coordinate systems $S'$, $S$ in a relative motion in $x$-direction with the speed $v=c\beta$. Clocks in both systems are synchronized at 
$t'=t=0$ when origins $O'$ and $O$ coincide. The problem was first time formulated by Einstein in 1905 in his revolutionary work\cite{Einstein} in the following way (with some changes in denotations): 

{\it ``At the time $t'=t=0$, when the origin of the coordinates is common to the two systems, let a spherical wave be emitted therefrom, and be propagated with the velocity $c$ in system $S$. If $(x,\ y,\ z)$ be a point just attained by this wave, then $x^2+y^2+z^2=c^2t^2$. Transforming this equation with the aid of our equations of transformation }\ (the LT is meant, a.v.) \ {\it we obtain after simple calculations \ $x'^2+y'^2+z'^2=c^2t'^2$.\ The wave under consideration is therefore no less a spherical wave with velocity of propagation $c$ when viewed in the moving system} \ ($S'$ is meant, a.v.).  {\it This shows that our two fundamental principles are compatible.'' }

In the above paragraph, the statement that the null quadratic metric form
\begin{equation}
(ct)^2-(x^2+y^2+z^2)=(ct')^2-(x'^2+y'^2+z'^2)=0 
\label{9}
\end{equation}
is a consequence of two SR postulates is true. There is the second statement about a shape of the wave front in a moving system; it is asserted that if the front originated at the $O$-event is spherical in one coordinate system then it is spherical in the other coordinate system. Notice, that Einstein did not indicate whether the source of flash is attached to any of coordinate systems. However, from the proposition that the {\it spherical} front is emitted, say, in $S$, it should be understood that the source is attached to the origin $O$ in $S$. The question arises if the picture of the {\it spherical} front originated in $S$ but observed in $S'$ is, indeed, predicted in SRT. We shall return to this question after a brief review of some references concerning the issue.
   
Pauli\cite{Pauli} was a student in 1921 when he wrote a book-review ``Theory of Relativity''. In the book he noted a paradox in the statement about two light spheres from the same source but viewed from different reference frames in relative motion: in his view, those spheres must be {\it different} that would be a contradiction. Nevertheless, he justified the two sphere picture and
concluded: {\it ``This contradiction disappears, however, if one admits that space points which are reached simultaneously for $O$, are not reached simultaneously for $O'$ ''}, (p.9). Descriptions of this issue in textbooks later on were more or less a repetition of Einstein's or Pauli's version, as in following examples.    
According to Reitz\cite{Reitz} (p.563), if {\it ``a light source at the common origin emits a light pulse''}, then both observers see a spherical wave front propagating outward (the state of motion of the source is not defined). There is no clue there about which frame the source is attached to. In Jackson's textbook\cite{Jackson} (p. 524) a source is specified. It produces a light flash from $O$ at $t=0$ when two origins coincide; the conclusion was made that {\it ``Einstein's second postulate implies that observers in  both $S$ and $S'$ will see a spherical shell of radiation expanding outward from the respective origins with speed $c$''}, (equation (\ref{9}) is referred).
In Tipler's textbook\cite{Tipler} (p.21): {\it ``A flash of light starts from the origin of $S$ at $t=0$. Since we have assumed that the origins coincide at $t=t'=0$, the flash also starts at the origin of $S'$ at $t'=0$. The flash expands from both origins as a spherical wave, in accordance with two postulates''}. 
In Panofsky's\cite{Panofsky} (p.243), the issue is discussed closely to Pauli's text.
In Rindler's\cite{Rindler} (p.75): {\it A moving sphere presents a circular outline to all observers...}.

There are topical questions in this issue need to be discussed next: a specification of a rest frame for the source of light; a light front shape seen from a moving frame; simultaneity of events involving photons.

Why  a specification of a rest frame for the source of light is important? A photon vector (just as a particle vector) is subject to the LT without inquiring where the particle or the photon came from. To establish a relationship of proper/improper quantities, one should specify the rest frame for the particle. The photon cannot be at rest, so the proper/improper categorization is not relevant to it. However, when the question arises about photon properties depending on emission versus observation angle with a source and a detector being in a relative motion, the source specification is important, as in problems of ray aberration and Doppler effects. Problems of isotropy and shape of light front observed in different reference frames and simultaneity of events involving photons also require a source specification.
In most references, as in the above and others relevant to the problems, a rest frame for the source was explicitly or indirectly specified; otherwise, a problem formulation would be underdefined (ill-posed). If defined, one needs a pen and a piece of paper to calculate a result with the use of LT technique, as in the following calculation of the shape of light front emitted from the source given.

\subsection{The shape of wave front in a moving system}

Let us put a light source at the coordinate origin $O'$ in John's frame $S'$ and a light detector at $O$ in Mary's frame $S$, both frames being in a standard geometrical configuration. It is assumed that at the instant 
$t'=t=0$ of origin coincidence (the $0$-event), both observers verified their clock synchronization. At the same instant, John triggered an {\it isotropic} light flash, which would reached a spherical surface of a radius $r'_0=c\Delta t'_0$. Obviously, the spherical surface $r'=r'_0$ in John's frame represent simultaneous events involving $O'$ photons from a ray of any polar angle of emission $\theta'$ in 
$x',\ y'$ plane. Those were physical events detectable by any observer, Mary in 
$S$, in particular. A ray of angle $\theta'$ in John's frame appears to her at some different angle $\theta$ dependent on $\beta$. This is the aberration phenomena known in pre-relativistic and relativistic theories, and it affects the shape of John's light front seen from different inertial frames. We are in position to calculate the shape of front in $S$.

Let us consider a 4-coordinate vector of the photon from a specified source in John's frame 
\begin{equation}
x'^\mu /c=(\Delta t'_0,\ \Delta t'_0 \cos\theta',\ \Delta t'_0 \sin\theta',\ 0)
\label{10}
\end{equation}
where $\Delta t'_0$ is a fixed time interval. The corresponding vector in Mary's frame is obtained from (\ref{10}) by the inverse LT, giving components of $x^\mu/c=(\Delta t,\ \Delta t \cos\theta,\ \Delta t \sin\theta,\ 0)$:  
\begin{equation}
\Delta t=\gamma \Delta t'_0 (1+\beta\cos\theta'), \ \ \Delta t\cos\theta= \gamma t'_0 (\cos\theta'+\beta),\  \ 
\Delta t\sin\theta=\Delta t'_0\sin\theta')
\label{11}
\end{equation}
from which, after simple algebraic calculations, one gets
\begin{equation}
\Delta t/\Delta t'_0=\gamma(1+\cos\theta')=1/\gamma(1-\beta\cos\theta)
\label{12}
\end{equation}
\begin{equation}
\tan\theta=\sin\theta'/\gamma(\cos\theta+\beta),\  \ \cos\theta=(\cos\theta'+\beta)/(1+\beta\cos\theta')
\label{13}
\end{equation}
\begin{equation}
\tan\theta'=\sin\theta/\gamma(\cos\theta-\beta),\  \ \cos\theta'=(\cos\theta-\beta)/(1-\beta\cos\theta)
\label{14}
\end{equation}
One can put for simplicity $c \Delta t'_0=1$, then, due to constancy of the speed of light
\begin{equation}
r(\theta)=c \Delta t(\theta)= \gamma(1+\cos\theta')=1/\gamma(1-\beta\cos\theta')
\label{15}
\end{equation}
This is a radius measured by Mary in $S$ when the source is attached to John's $S'$ frame. The radius is a function of a polar emission angle $\theta$ in $S$ corresponding to $\theta'$ in $S'$ . Both angles are connected by aberration formulae (\ref{13}), (\ref{14}). Clearly, Mary sees an ellipse (\ref{15}) in $(x,\ y)$ plane, or an elongated ellipsoid in 3-space with semimajor and semiminor axes $a=\gamma$ and $b=1$, correspondingly, and the eccentricity $\epsilon=\beta$. 
The ellipse is shown in Fig \ref{Ellipse} for $r'_0=c\Delta t'_0=1$ and $\beta=0.7$. It is similar to that presented in Vankov\cite{Vankov} for photon 4-momentum vectors (the similarity is a consequence of complementarity of coordinate and momentum space).

\begin{figure}[t]
\includegraphics{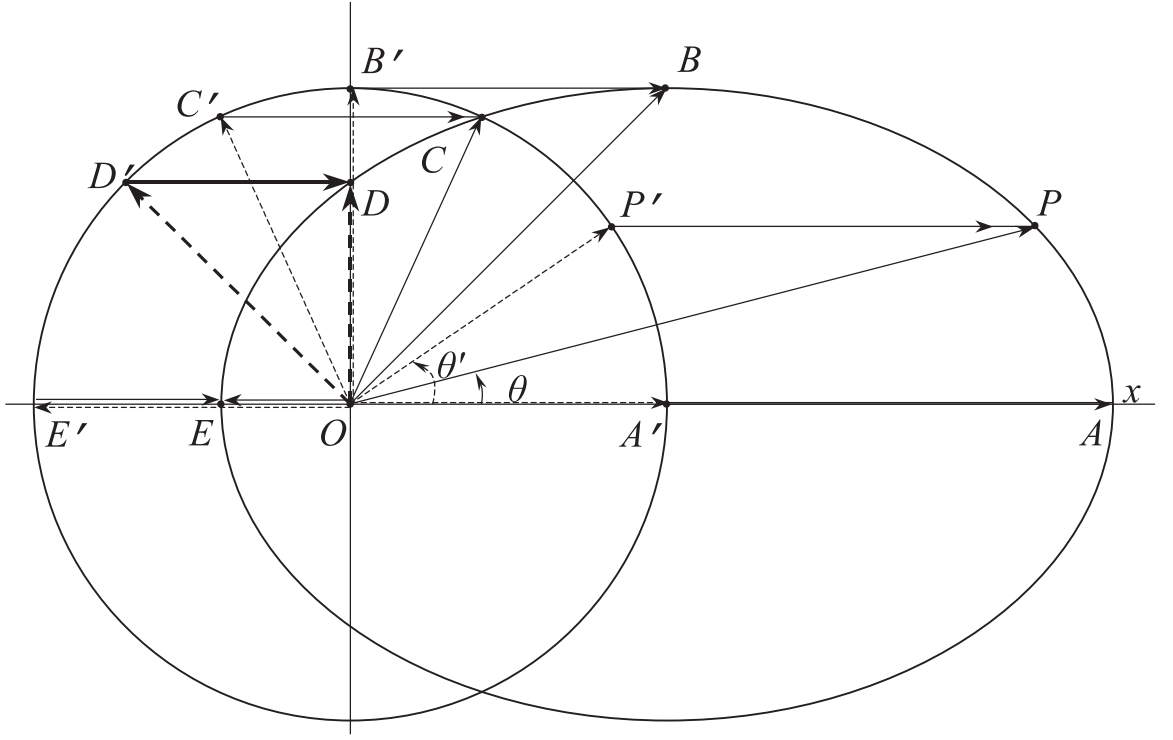}
\label{Ellipse}
\caption{\label{Ellipse} Sphere-to-ellipsoid conversion in coordinate transformations}
\end{figure}

Back to sphere-versus-ellipsoid controversies in literature, as reviewed above: evidently, authors did not pay much attention to the source issue, and did not
attempt to calculate. Hypnotized by (\ref{9}), they jumped to the wrong conclusion that if one observer saw an isotropic radiation (a sphere) than all observers from different inertial frames would see the same picture, provided the initial condition $t'=t=0$ for a light flash was arranged. In our calculation, the source was put at the origin of $S'$ (John's) coordinate system. Therefore, $r'=r'_0=\sqrt{(x'^2+y'^2+z'^2)}=c\Delta t' $ is a sphere of isotropic radiation. At the same time,  $r=r(\theta)=\sqrt{(x^2+y^2+z^2)}=c\Delta t (\theta)$ in $S$ (Mary) is an ellipsoid of anisotropic radiation containing the same photons, as on John's sphere. The LT realizes sphere-to-ellipsoid mapping, which results in temporal and spatial (radial) rescaling by the same factor $1/\gamma(1-\beta\cos\theta)$, a function of a polar angle $\theta$ with the boost parameter $\beta$. Recall that relativistic space-time rescaling means a change of length-time units due to a relative motion in a given direction. 
Events at the ellipsoid 4-point (except for symmetric ones for any pair of angles $\pm\theta$) are, of course, not simultaneous. 

Relativistic Physics of the aberration and Doppler phenomena is not complete without a consideration of the momentum transfer due to ``vector boost'' and the corresponding ellipsoid picture in the 4-momentum (complementary) space \cite{Vankov}. In this connection, there is an interesting
relationship between $|\overrightarrow {OP}|=r$ and $|\overrightarrow {OP'}|=r'_0=1$. The vector diagram reads  $\overrightarrow {OP}= \overrightarrow {OP'}+ \overrightarrow {P'P}$ where  
$|\overrightarrow {P'P}|=\Delta x$ is a magnitude of a boost vector in $x$-direction in the LT diagram for photon 4-vectors. The addition is analogous to the momentum transfer in the 4-momentum space.
As a LT consequence, $\Delta x(\theta)$ as well as 
$r(\theta)$ are linear functions of $\cos\theta'$    
\begin{eqnarray}
\Delta x=(\gamma-1)\cos\theta'+\gamma\beta, \  & \ r=\gamma +\gamma\beta\cos\theta'
\label{24}
\end{eqnarray}
In a low-speed approximation, $\Delta x=\beta$,  \ $r=1+\beta\cos\theta'$. 

Several cases illustrating the boost for different angles are shown in the picture and commented below.

{\it Case A}. \ At $\cos\theta^\prime =\cos\theta =1$  $r=\gamma(1+\beta)=\sqrt{(1+\beta)/(1-\beta)}$;\ \ $\overrightarrow {OA}= \overrightarrow {OA'}+ \overrightarrow {A'A}$ ; \ $\Delta x=(\gamma -1)+\gamma\beta$. 
If a frequency is measured, a maximal blue-shift will be determined.

{\it Case B}. \  At $\cos\theta^\prime =0$,\ $\cos\theta =\beta$, $r=\gamma$ \ $\overrightarrow {OB}= \overrightarrow {OB'}+ \overrightarrow {B'B}$; \ $\Delta x=\gamma\beta$.
In this case, a ray perpendicular to the $x'$-axis in John's frame is emitted. From the time interval measured, the time dilation effect will be determined. 
 
{\it Case C}. \ At $\cos\theta^\prime =-(\gamma-1)/\gamma\beta$,\ $\cos\theta =(\gamma-1)/\gamma\beta$\ ,  
$r=1$. \ $\overrightarrow {OC}= \overrightarrow {OC'}+ \overrightarrow {C'C}$, \  $\Delta x=2(\gamma-1)/\gamma\beta) $. 
If a frequency is measured, no shift is found.

{\it Case D} \ At $\cos\theta^\prime =-\beta$, $\cos\theta =0$,\ \ $r=1/\gamma$\ \ $\overrightarrow {OD}= \overrightarrow {OD'}+ \overrightarrow {D'D}$,\  $\Delta x=\beta $. 
In this case, a ray perpendicular to the 
$x$-axis in Mary's frame is detected. If a frequency is measured,
the transverse Doppler (red-shift) effect will be determined equivalent to the time-dilation effect.

{\it Case E}. \ At $\cos\theta^\prime =\cos\theta =-1$, \ $r=\gamma(1-\beta)=\sqrt{(1-\beta)/(1+\beta)}$,\  $\overrightarrow {OE}= \overrightarrow {OE'}+ \overrightarrow {E'E}$ , \ $\Delta x=-(\gamma -1)+\gamma\beta$. 
If a frequency is measured, a maximal red-shift will be found. 

It is interesting to notice that for every ray of the directional cosine $\cos\theta^\prime$ in the primed coordinate system another ray of the directional cosine  $\cos\theta$ exists in the unprimed system such that $r(\cos\theta^\prime)\cdot r(\cos(\pi-\theta))=1$ identically, as in A and E cases and B and D cases. The case C is unique: there is no temporal and spatial rescaling for the directional cosine  $\cos\theta^\prime =-(\gamma-1)/\gamma\beta$. A continual deviation from this direction leads to either contraction ($r<1$) or extention 
($r>1$) of time/length scale up to maximal effects at  $\cos\theta^\prime=\pm 1$. In particular, one can verify the identity $r(0)\cdot r(\pi)=1$.

In our view, a historical  ``picture of two spheres'' has made pedagogical and ideological harm and was a source of many confusions. Let us see some examples. The paper\cite{Scherr} from the Physics Educational Research AJP Section presents the problem in a worst (ill-posed) variant, when a relative motion of a source with respect to an observer is not specified, and the wrong answer is stated to be taken for granted. In the next quotation from the paper, we changed names of imaginary  observers for convenience. 

{\it ``Students are told that two observers, John and Mary, move past each other at relativistic relative speed. At the instant they pass, a spark occurs between them, emitting a flash of light''}.

Given this, students  {\it ``are asked to illustrate the fact that the speed of light is the same in all directions according to John''}. 

Then, students {\it ``need to recognize that Mary also observes the propagation of light to be isotropic. Thus, she is at the center of a spherical wave front in her frame... This exercise is not difficult for most students. However, it lays important groundwork for the subsequent exercise (train paradox'')}. 

In physical reality, Mary sees an ellipsoid of anisotropic radiation. One could imagine what a mess could be built up in students' heads in a course of ``guiding'' (brain-washing) exercises composed to teach wrong concepts.

As an example of ideological controversy, a work\cite{Pierseaux} is worth mentioning, in which {\it ``Einstein's spherical waves versus Poincare's ellipsoidal waves''} is discussed. It reveals the fact that
Poincare in early 1900's suggested an elliptic picture  (though in the methodology of ether) of a light front seen by a moving observer. The author\cite{Pierseaux} concluded that both pictures are logically justified: Einstein's one is consequence of his ``convention of clock synchronization'', while Poincare's  elliptic picture is ``a direct geometrical representation of Poincare's simultaneity'' in the ether concept.
In this connection, it is stated that known (Einstein's) relativistic Doppler formula ``presents a strange aspect'';  ``another relativistic Doppler formula'' is suggested, both formulae to be subject to experimental retesting. In our view, this adds unnecessary controversy in literature. The truth is that the problem of shape of light front is, indeed, tightly related to the aberration and Doppler effects but there is no physical motive to revise the Doppler effect on this ground. The formulae (\ref{12}-\ref{15}) have no alternatives unless the SRT is claimed to be subject to radical revision either due to incompetence or an attempt of revival 
of the ether concept at a new (cosmological) level. This issue was discussed earlier\cite{Vankov}.

Why not to experimentally check the ellipsoid-versus-sphere picture? To answer this question, one has to take into account the fact that the aberration and Doppler effects, correspondingly, have leading first-order in $\beta$ terms what means that the effects are observed in pre-relativistic physics consistently with slow-motion observations to the precision of second order terms. In this approximation, the ellipsoid (\ref{15}) becomes a sphere around a center shifted from the origin to a point 
$\beta$ (but the observational radius $r(\theta)$ is kept originating at the common origin).
In other words, ellipsoid-versus-sphere difference is a true relativistic (of a second order) effect.
The corresponding experimental test of high precision would be costly and should be motivated by something more substantial than an individual confusion or textbook controversies. 

In the previous Section, well known time dilation effect independent of direction was discussed. In our view, standard relativistic effects, such as time dilation and length contraction, should be considered a particular case of a general properties of space-time transformation in Minkowski space.
The ellipsoidal shape of a wave front of light demonstrates space-time rescaling effects resulting from the LT of null vectors. As an additional exercise, one can study space-time rescaling arising under general Lorentz transformations of time-like vectors.

\subsection{Einstein's train/embankment experiment and relativity of simultaneity}

The problem illustrating the relativity of simultaneity may be formulated in terms of SRT, as follows. A train (the $S'$ frame) is characterized by a rest length interval $\Delta x'_0$ between end point $x'=A'=0$ and $x'=B'=\Delta x'_0$, and it moves uniformly at the speed 
$v=c\beta$ in $x$-direction with respect to the embankment (the $S$-frame). Two points, $x=A=0$ and $x=B=\Delta x_0$, on the embankment 
indicate a similar rest interval $\Delta x_0$, that is $\Delta x'_0=\Delta x_0$ by definition. It is assumed that two lightnings strike embankment points $A$ and $B$ in $S$ {\it simultaneously} at $t=0$.
According to Einstein's simultaneity criterion, two lightnings having stricken the end points at the instant are simultaneous, if photons from flashes at each end intersect at midpoint $M$ between $A$ and $B$.
Will same photons reach midpoint $M'$ between $A'$ and $B'$ in $S'$ {\it simultaneously} if points $A'$ and $A$ coincide at the moment of strikes $t'=t=0$? 

This problem has a simple solution in SRT if one takes into account that the $0$-event specified. Two simultaneous events (two flashes) in one coordinate system ($S$) are not simultaneous in another coordinate system ($S'$) in a relative motion: the difference of photon arrivals to the midpoint $M'$ is $t_A'-t_B'=\gamma\beta \Delta x_0/c$. In the so-called pre-relativistic (low speed) approximation, it is  $t_A'-t_B'=\beta \Delta x_0/c$. It is seen that a non-simultaneity time difference is a linear in $\beta$ effect, that is why it does not vanish in pre-relativistic photon kinematics.

The famous thought (train/embankment) experiment suggested by Einstein\cite{Bergman,Einstein2} was intended to illustrate the relativity of simultaneity {\it prior to student knowledge of SRT}. So, the problem was actually formulated in the pre-relativistic approximation of photon kinematics, in which a photon would propagate with the speed $c$ {\it in a particular reference frame} while the Galilean relativity having been respected but the postulate of the speed of light constancy was not introduced. Consequently, two length intervals, which are equal at rest, remain equal in relative motion (probably, for this reason, Einstein did not distinguish between {\it coinciding} points $A'$ and $A$ or $B'$ and $B$, and used common $A$ and common $B$ points). Einstein did not commented his simplification of the train/embankment experiment, and did not specified frames, in which sources of two flashes were at rest.  However, from his statement that photons reached midpoint simultaneously in the embankment frame, one should guess that sources were at rest with respect to the embankment. From a simple deliberation of the picture of photons flying either in a positive or negative $x$ direction with a relative speed $c+v$ or $c-v$, correspondingly, with respect to the train, the non-simultaneity effect described by the formula  $t_B-t_A=\beta \Delta x_0/c$ (which is the low-speed SRT approximation) follows. One can immediately figure out that photons from two flashes intersect the point $X'$ between $A'$ and $B'$ in the proportion $A'X'/X'B'=(1-\beta)/(1+\beta)$. Photons will reach the point $X'$ in the train frame simultaneously is correct (in the slow motion approximation of Relativity Theory). In fact, Einstein proposed to check his definition of simultaneity, which is obvious for two spatial points at rest but became not applicable in the case of relative motion.

It should be noted that the conclusion drawn from the thought experiment is in contradiction with ``the two-sphere picture'' previously discussed.
Let us consider the train/embankment pre-relativistic scenario with $S'$ and $S$ observers at end points coinciding at $t'=t=0$ when flashes occured at $A$ and $B$. According to ``the two-sphere picture'', each observer at $A'$ and $B'$ should see 
a sphere, which expands {\it isotropically with the speed of light}. Hence, the observers would witness that two spherical fronts starting simultaneously from $A$ and $B$ touched each other at the midpoint (but not at $X'$ point).

In the cited ``guiding exercises''\cite{Scherr}, the Einstein's result of the train/embankment problem was stated on the wrong premises such as: a) the propagation of light is isotropic in all frames; b) the relativity of simultaneity is a consequence of the invariance of the speed of light. This is a misleading presentation of the famous Einstein's thought experiment.

Einstein's idea to elucidate the relativity of simultaneity in the train/embankment scenario prior to knowledge of Relativity Theory was logically acceptable and technically reasonable at the time of theory popularization. Classical theory of light aberration and Doppler shift in vacuum was known well before Einstein's postulate of speed of light constancy, and the corresponding (first-order) effects were measured in early 1900s. Having this, the (first-order) effect of frame dependence of simultaneity is, indeed, understandable in a fashion of classical light theory. From this historical viewpoint we cannot agree with Nelson's criticism\cite{Nelson} of Einstein's train/embankment experiment. The criticism is solely focused on the alleged conflict of the Einstein's scenario with the postulate of speed of light constancy. But there is no such a conflict at all because Einstein had good reasons not to care about the postulate (which is violated in second-order terms in pre-relativistic models). Ironically, the Nelson's claim was immediately denied by ``textbook advocates''\cite{Mallinckrodt} who argued that Nelson was thrice wrong in his criticism because {\it Einstein in his scenario applied the postulate not once but twice} right away... 

The next section is devoted to controversies related to problems, which were earlier discussed in the context of complementarity of coordinate and momentum 4-spaces\cite{Vankov}. 

\section{ Does a moving clock run slower? The clock (twin) paradox.}

The content of this Section contains interrelated topics of the SRT space-time rescaling phenomenon, which were
left blank in previous Sections: 

-  significance and operational meaning of the $0$-event; 

-  physical treatment of the time dilation effect; 

-  the clock (twin) paradox.

\subsection{The role of the $0$-event in Lorentz transformations}

The LT manifests Minkowskian metric invariance through proper/improper relationship of time and length units.  The corresponding measuring procedures were described in the Section 2 with the exception of an initial clock synchronization, the $O$-event. It is assumed that each observer in $S$ and $S'$ carries a standard clock located in the coordinate origin of each coordinate system. The clock can play a role of a light signal emitter or a corresponding detector. Thus, observers can exchange observational data at the instant $t'=t=0$ of origin coincidence (the $0$-event); they conduct a similar procedure at any other event  labeled $P$ in $S$ or $P'$ in $S'$. 
The direct (or inverse) LT, when applied to a 4-vector $\overrightarrow {OP}$ (or $\overrightarrow {OP'}$), establishes a relationship between points $P$ and $P'$. To make it possible, it is required in the LT concept that the $0$-event is operationally realizable and computationally fixed. 

Thus, in the Minkowski metric determination, the $0$-event is an arbitrarily chosen common zero 4-point, from which a 4-distance is measured to any point (event) $P$ in any coordinate system. In our example, the $0$-event was an emission of light signal by John's clock in $S'$ at $t'=\Delta t'_0$ in $y$-direction and a detection of the signal by Mary's detector in $S$. Strictly speaking, two physically different events occur (an emission and a detection) at any event $P$ in a process of information exchange between two imaginary observers. 

When we say that two coordinate systems are in relative motion in $x$ direction, it is often meant that axes 
$x$ and $x'$ coincide. If so, both clocks will collide at the initial moment. To avoid this illogical situation, one should allow some distance between $x$ and $x'$ axes. Let us assume that one observer (say, John in $S'$) sends a short light signal in $y$-direction while the other observer (say, Mary in $S$) keeps her detector on to detect the signal at an instant John passes her. Both agreed that it would be the instant 
$t'=t=0$ at $0$-event in spite of some physical delay of a signal propagation. 
In general, to detect any event $P(t,\ x,\ y,\ z)$, observers need a multitude of clocks in their possession, potentially, in every point of 3-space in $S$ and $S'$, respectively.

In our example, John creates an original event $P'$, and Mary observers its image $P$. From (\ref{12}), it is seen that a light signal emitted by John at $t'=\Delta t'_0$ perpendicularly to the direction of motion, that is $\cos{\theta'}=0$, will be detected by Mary at
$t=\Delta t=\gamma\Delta t'_0$ by the clock-detector located at $x/c=\beta\Delta t$. The angle of observation $\theta$ is not $\pi/2$ because of aberration effect. or ssimplicity, one may take $\Delta t'_0=1$ $s$.  Then, the measured by Mary time interval will be $\Delta t > 1$ $s$ as far as $\gamma>1$.

\subsection{On the time dilation effect}

The time dilation effect $\Delta t=\gamma\Delta t'_0$ follows from (\ref{12}). This is the case of the $\beta$-boost when the light signal emitted at $t'=\Delta t'_0$ perpendicularly to the direction of relative motion in $S'$ is detected in $S$. When we derived the time dilation effect by applying the LT to the 4-coordinate vector of particle, we meant the procedure of information exchange by means of light signal, but did not get into such concrete details. 

There is a textbook cliche expressing a physical meaning of the effect: ``a moving clock runs slower''.  Such a clock does not exist, however, because the improper time interval is measured by, at least, two clocks located at different places along a direction of motion. In reality, standard clocks in all inertial reference frames run identically in terms of their own (proper) time records, and a comparison procedure is suggested in terms of proper versus improper records. It would be illogical to expect that two initially synchronized clocks in a relative motion could show different proper time records when brought back to the initial point, provided possible effects of non-inertial stages of motion were eliminated. Before discussing this issue (the so-called clock paradox), let us consider the time dilation effect in connection with an ellipsoidal picture of light front in a moving frame. The picture illustrates relativistic aberration (\ref{13}, {14}) and Doppler effect (\ref{12}) as a result of time unit rescaling. Suppose, observers conduct the measurements periodically with the frequency $f'_0=1/\Delta t'_0$\ ({\it cycle per second}). The light frequency from a moving source is 
\begin{equation}
f(\theta)=f_0/\gamma (1-\beta\cos\theta).
\label{16}
\end{equation}
Thus, it is clear that the transverse Doppler effect is directly related to the independent of direction time dilation effect in the case, when the LT is applied to the coordinate vector aligned in parallel with a direction of motion (1D problem). In general, the time unit rescaling factor depends on an angle $\theta$, as in the sphere versus ellipsoid picture (3D problem).

\subsection{The so-called clock paradox}

Recall that paradox is about a comparison of time records of two initially synchronized clocks, one being at rest while the other traveled at a constant speed from $A$ to $B$ and back. When they met, their time record should be the same due to relativity of motion (provided the effect of a change of direction was eliminated). On the other hand, the $A$ observer (say, Mary) had to register a round trip time $2t_0$ ($t_0$ is a one-way time) using her wristwatch, while her traveling counterpart (say, John) carried his wristwatch in motion with respect to $A$. Hence, in accordance with the the time dilation concept, his clock run slower; specifically, it would show lesser time record  $2t_0/\gamma $. This is a typical textbook (alleged) solution considered as a well established fact.
We analyzed the paradox in detail previously\cite{Vankov}; the main arguments are summarized here. 

There is no paradox because an apparent contradiction is resolved. The problem is that opposite solutions were given. As opposed to textbooks, we concluded that no difference in clock records would be shown. 
In our view, the textbook solution is incorrect because it came from a) the ill-posed problem formulation (in part of end point conditions), and b) a wrong interpretation of time dilation concept (when a procedure of clock record comparison is considered). The solution should be based on calculations by a LT means along with a correct physical interpretation. Let a distance between $A$ and $B$ be $x_0$: this is the proper distance for Mary in terms of her standard (unit) measuring rods. She measures a one-way travel time by a time-of-flight method: $x_0=c\beta t$, where 
$t^{(1)}=t_B-t_A$ is the improper time obtained as a difference of clock records $t_A$ at the starting point 
$A$ (the $0$-event in the first half of travel), and $t_B$ at the end point $B$. The return trip should be considered separately and symmetrically: $t^{(2)}=t_A-t_B$ with a clock record $t_B$ at the starting point 
$B$ (the $0$-event in the second half), and a clock record $t_A$ at the end point $A$. All clocks there are not wristwatches, they are Mary's clocks placed off the $x$-axis (side clocks). The results of Mary's measurements should be a total (improper) time of John's travel $2t=t_A+t_B$ and a Doppler-shifted frequency of John's signal $f$. It is easy to prove that the product $2t\cdot f=2n_0$ is a number of clock ticks registered by each observer by his/her wristwatch; this number is Lorentz invariant. In view of a long history of debates around the paradox, its current (controversial) status in literature, and beliefs among physical community, we consider the paradox a great historical myth in Physics. 

\section{Variable proper mass concept in SRT Dynamics}

SRT Kinematics follows from SRT Dynamics when a field of forces vanishes. However, there is no unique formulation of SRT Dynamics\cite{Landau, Goldstein}. In particular, the concept of the proper mass could be different depending on the concept of the Minkowski force $K^\mu$ acting on a test particle of proper mass $m$. In GRT and conventional Relativistic Mechanics, the proper mass is assumed to be constant $m=m_0$, so the generalized  Newton's law has the form 
\begin{equation} 
dP^{\mu}/ds=d(m_0 u^\mu) /ds=m_0 du^\mu /ds= K^\mu 
\label{1r}
\end{equation}
Our viewpoint is that the proper mass constancy is a long-standing dogma in Physics. This assumption is neither justified by direct experiments nor follows from first physical principles: its revision is the issue of theory physical foundations and subject to experimental falsification. The importance of the proper mass problem is seen, in particular, in the quantum connection of mass and time $mc^2=hf$, where $h$ is the Plank constant and $f=1/\Delta t$, and the de Broglie wave relations. It shows a dependence of a clock rate on field strength.

Let us consider consequences of changing the assumption and allowing the proper mass to be field-dependent $m=m(s)$:
\begin{equation} 
K^\mu=dp^\mu/ds=u^{\mu}(dm/ds)+m(du^{\mu}/ds) 
\label{2r}
\end{equation}
Now we have a non-zero tangent component of Minkowski force $u^{\mu}(dm/ds)$. It should be noted that the proper mass variability is not a new idea: it was discussed in classical books by Synge \cite {Synge} and Moller \cite {Moller} but did not draw much attention among physics community. We are going to show that the introduction of the field-dependent proper mass in the relativistic Lagrangean framework is consistent with (\ref{2r}). Further, the results\cite{Vankov2} are reproduced in a more consistent form.

In order to derive equations of motion by applying the Hamilton's extremal action principle, let us introduce the Lagrangian for a particle in a potential field due to the source of mass $M>>m_0$
\begin{equation} 
L(s)=m(s)+W(s)
\label{3r}
\end{equation}
where $s=s(x^\mu)$ is a world line (arc)length, and a field is characterized by potential energy $W(s)$ for a repulsive force; it would be negative for an attractive force. The concept of potential energy requires that $W(x^\mu)\to 0$ and $m(s)\to m_0$ at $x^\mu \to \infty$ (an equivalent mass-energy unit is used for convenience). 

The Lagrangian $L(s)$ in (\ref{3r}) is used in a variation procedure for the extremal action $S$
\begin{equation} 
\delta S=\delta \int_a^b L(s)ds=\int_a^b \delta (L u^\mu )dx_\mu=0 
\label{5r}
\end{equation}
Thereafter, $m(s)$, $W(s)$ and $ds$ are subject to variation, bearing in mind 
that $s=s(x^\mu)$, 
$ds=\sqrt{ds^2}=\sqrt{dx^\mu dx_\mu}$, and also $dx^\mu/ds=ds/dx_\mu=u^\mu$.

Components of 4-position vector $x^\mu$ comprise a set of dynamical variables. They indicate 4 degrees of freedom of a test particle in a field. Thus, $ds$ is a subject to variation due to independent variations $\delta x^\mu$. In this sense, a varying proper mass $m(s)$ of a test particle should not be considered an additional dynamical variable manifesting the fifth degree of freedom of the system. The proper mass is determined by the potential and characterizes the binding energy $m(s)-m_0$. So, both $m(s)$ and $W(s)$ are put on the same footing.

From (\ref{5r}) to continue
\begin{equation} 
\delta S= \int_a^b \frac {\partial (L u^\mu)}{\partial x^\nu} \delta x_\nu dx_\mu=0
\label{6r}
\end{equation}
Because variations $\delta x_\nu$ are independent for different $\nu$, the equality $\delta S=0$ in (\ref{6r}) is possible if and only if 
\begin{equation}
\frac{\partial \left[L(s)u^\mu(s)\right]}{\partial x^\nu}=0
\label{11r}
\end{equation}
or equivalently
\begin{equation}
\frac{\partial \left[L(s)u^\mu(s)\right]}{\partial s} =0
\label{11rr}
\end{equation}
Finally, with the Lagrangian (\ref{3r}) substituted into (\ref{11rr}), we have the equations of motion
\begin{equation}
\frac{\partial \left[m(s)u^\mu(s)\right]}{\partial s}=-\frac{\partial \left[W(s)u^\mu(s)\right]}{\partial s}
\label{12r}
\end{equation}
Together with the equation 
\begin{eqnarray}
u_\mu u^\mu=1, \ & \ u_\mu (du^\mu/ds)=0
\label{13r}
\end{eqnarray} 
characterizing the time-like character of massive particles, they allow us to determine five correlated quantities $x^\mu(s)$, $m(s)$. 
The case of Lagrangian $L(s)=-m_0$ (a free particle)  for the action variation 
\begin{equation}
\delta S=m_0 \delta \int_a^b ds=m_0 \int_a^b d(\delta s)=0          
\label{7rrr}
\end{equation}
follows from (\ref{6r}).

The equations (\ref{12r}) contain two orthogonal Minkowski force 4-vectors, one acting along the world line (the tangent, or parallel component) and the other (orthogonal) acting perpendicularly to the world line: 
\begin{eqnarray} 
{\bf u} (dm/ds)={\bf K}_\parallel \ , \   m(d{\bf u}/ds)={\bf K}_\perp, \ & ({\bf K}_\parallel \cdot {\bf K}_\perp) =0 
\label{14r}
\end{eqnarray}
where \  $u^\mu (dm/ds)=K_\parallel^\mu= -u^\mu {\partial W}/{\partial s}$, \ \ $dm/ds=K^s=K_\parallel$ ,  \ \  $m (du^\mu/ds)= K_\perp^\mu =-W (du^\mu/ds)$.
So the equations (\ref{12r}) can be expressed in the convenient 4-vector form
\begin{eqnarray} 
d(m {\bf u})ds=d{\bf P}/ds={\bf K}_\parallel +{\bf K}_\perp ={\bf K}
\label{15r}
\end{eqnarray}
or
\begin{eqnarray} 
d(mu^\mu)ds=dP^\mu/ds=K^\mu
\label{16r}
\end{eqnarray}
 
Under weak-field conditions, predictions of the SRT Mechanics with a field-dependent proper mass is hard to distinguish from corresponding conventional results. Deviations rise with field strength, as discussed in our work\cite{Vankov2}. One of the remarkable results is an elimination of a classical problem of self-energy divergence for $1/r$-potential field.

\section{Summary and conclusions}

Relativistic concepts of aberration and Doppler effect, as well as relativity of simultaneity, are related to effects, which are described in SRT Kinematics of photon by formulae with leading terms linear in $\beta$. Not surprisingly, the aberration and Doppler phenomena were well understood before the SRT advent in terms of pre-relativistic photon model. In Galilean Kinematics of particles, the speed of light was assumed to be infinite. Consequently, Newtonian Physics dealt with absolute space and time and events, which could occur simultaneously regardless of reference frame choice. However, in pre-relativistic photon concept, photons propagate in space at the speed $c$ with respect to a source (but not a detector). In this concept, the Galilean relativity principle is respected but the invariance of $c$ is violated in the second and higher order terms. To this precision, the Einstein's idea of train/embankment experiment retrospectively was a good illustration of relativity of simultaneity. The controversy about it in literature causes many confusions.

Einstein's Special Relativity Theory made a revolutionary impact on further Physics development. Introduction of the postulate of speed of light constancy in addition to the Galilean principle of relativity of motion made time and length units dependent on motional state of observer (space-time rescaling). Consequently, proper versus improper physical quantities were introduced, in particularly, the improper time and improper length related to the so-called time dilation and length contraction effects. Those are relativistic effects determined by the second and higher order terms. At small $\beta$, the approximation  $\beta^2\to 0$, $\gamma\to 1$ leads to the Lorentz-to-Galilean transformation reduction and pre-relativistic photon Kinematics. Thus, SRT principles are clear and self-consistent. However, controversies in textbooks remain; they concern SRT applications and physical interpretation of the relativistic effects. We conducted a critical (retrospective) analysis of some controversies related to the pre-relativistic photon Kinematics as well as relativistic effects (the time dilation effect and its consequences). The conclusion was made that controversies caused by lack of rigor in some textbook presentations of relativistic effects make a negative influence on student Physics education. Many physicists and specialists in different branches of Engineering and Technology, probably, have already received ambiguous knowledge about Relativity Theory from ``popular'' presentation of such issues as relativity of simultaneity in Einstein train/embankment experiment, constancy of the speed of light from  viewpoint of observers in different inertial frames, time dilation effect sown by a clock running slower in motion, and others. 

In this paper, a rigor treatment of relativistic effects at two levels is attempted: a) for interested people (confused about Relativity Theory) whose specialities require General Physics knowledge;  b) for experts in fields of Modern Physics, especially, in relativistic theories who may disagree on or broaden some our points up to further competent disputes. In any case, this is the reader who decides.



\begin{thebibliography}{99}

\bibitem{Baierlein} R. Baierlein. ``Two myths about special relativity''. Am. J. Phys. {\em 74}(3), p. 193-195 (2006). Also: Private communication.
\bibitem{Peres} Asher Peres. ``Quantum Theory: Concepts and Methods''. Kluwer Academic Publishers, Dordrecht, Boston, London, (1993).
\bibitem {Sartori} L. Sartori. ``Understanding Relativity''. University of California Press, Berkeley, Los Angeles, London, (1984).
\bibitem{Einstein}  A. Einstein ``On the Electrodynamics of Moving Bodies''. Annalen der Physiks, {it 17} (1905), p. 891-921. Also in ``The Collected papers of Albert Einstein'',  John Stachel, Editor, Vol. 2. Princeton University Press (1989), p. 275-306. Also, ``The Principle of Relativity'', Introduction and comment by Sommerfeld, (English translation), Dover, New York, p. 37-65 (1952). 
\bibitem{Pauli} W. Pauli. ``Theory of Relativity''. Translated from German by G. Field. Dover Publications, Inc., New York (1981). Originally appeared in German ``Relativitatstheorie'', Encyklopadie der matematischen Wissenschaften, Vol. V19, B. G. Teubner, Leipzig,(1921).
\bibitem {Reitz} J. Reitz, F. Milford, R. Cristy. ``Foundations of Electromagnetic Theory''. Addison Wesley Longman (1993)
\bibitem{Jackson} J. Jackson, ``Classical Electrodynamics'', John Wesley and Sons, New York 1998) 
\bibitem{Tipler} P. Tipler, R. Llewwellyn. ``Modern Physics''. W. h. Freeman and Co. New York (1999).
\bibitem{Panofsky} W. Panofsky, M. Phillips. ``Classical Electromagnetic Theory''. Addison-Wesley Co (1955).
\bibitem{Rindler} W. Rindler.``Relativity: Special, General, and Cosmological''. Oxford Press (2001). Also:  W. Rindler. ``Essential Relativity''.  W. Rindler. Van Nostrand Reinhold Co. (1969).
\bibitem{Vankov} A. Vankov. arXiv: physics/0603168 ``Mass, time, and clock (twin) paradox in relativity theory''. 
\bibitem{Scherr} R. Scherr, P. Shaffer, and S. Vokos. ``The challenge of changing deeply held students beliefs about the relativity of simultaneity'', Am. J. Phys. {\it 70}, 12, p. 1238-1248 (2002).
\bibitem{Pierseaux} Y. Pierseaux. ``Special Relativity: Einstein's spherical waves versus Poincare's ellipsoidal waves'', arXiv:physics/0411045 (with references to Poincare's works). ``Einstein's relativistic Doppler formula'',  arXiv:physics/0509163. Also: Private communication.
\bibitem{Bergman} P. Bergman. ``Introduction to the Theory of Relativity'' with a Foreword by Albert Einstein 
(1942). 
\bibitem{Einstein2} A. Einstein. ``Relativity'' (authorized translation by Robert W. Lawson). Three River Press, New York (1961).
\bibitem{Nelson} A. Nelson. ``Reinterpreting the famous train/embankment experiment of relativity''. Eur. J. Phys. {\it 24}, (2003), p. 379-396. 
\bibitem{Mallinckrodt}. Eur. J. Phys. (Letters and Comments). {\it 25}, (2004): D. Rowland, L45-L47; Mallinckrodt, L49-L50; A. Nelson, L51-L56.
\bibitem {Landau} L. D. Landau and E M. Lifshitz.
``The Classical Theory of Fields'', Pergamon Press (1975).
\bibitem {Goldstein} H. Goldstein. ``Classical Mechanics''. Addison-Wesley Publishing Company; series in Advanced Physics (1950). 
H. Goldstein, C. Poole, J. Safko.  ``Classical Mechanics'', 3d Edition. Addison-Wesley Publishing Company (2001).
\bibitem {Synge} J. L. Synge. ``Relativity: The Special Theory''.
 North Holland Publishing Company, Amsterdam (1965). ``Relativity: The General Theory''.
 North Holland Publishing Company, Amsterdam (1964). 
\bibitem {Moller} C. Moller. ``The Theory of Relativity''. The international series of monographs on physics. Delhi, Oxford University Press (1972).
\bibitem {Vankov2} A. Vankov. ``On the de Broglie Wave Nature''. Annales de la Fondation Louis de Broglie. 
Vol. {\bf 30}, 1, page 15 (2005). 
A. Vankov. ``Testing Relativistic Mass-Energy Concept in Physics of Gravity and Electricity''. Annales de la Fondation Louis de Broglie. (The Special Issue dedicated to the 50th anniversary of the Yang-Mills, 1954, paper). Vol. {\bf 29}, 2, page 1035 (2004). 




\end{thebibliography}
\end{document}